\documentclass[prd,superscriptaddress,amsfonts,amssymb,amsmath,showpacs,twocolumn]{revtex4-2}
\usepackage{bm}
\usepackage{amsfonts}
\usepackage{latexsym}
\usepackage[latin1]{inputenc}
\usepackage{graphicx}
\usepackage{amsmath}
\usepackage{palatino}
\usepackage{mathpazo}
\usepackage{textcomp}
\usepackage{float}
\usepackage{booktabs}
\usepackage{dcolumn}
\usepackage{hyperref}
\hypersetup{colorlinks,citecolor=blue}
\usepackage{amsmath}
\usepackage{xcolor}
\usepackage{orcidlink}
\usepackage[caption=false]{subfig}
\usepackage{commath}
\def\jnl@style{\it}
\def\aaref@jnl#1{{\jnl@style#1}}

\def\aaref@jnl#1{{\jnl@style#1}}

\def\aj{\aaref@jnl{AJ}}                   
\def\apj{\aaref@jnl{ApJ}}                 
\def\apjl{\aaref@jnl{ApJ}}                
\def\apjs{\aaref@jnl{ApJS}}               
\def\apss{\aaref@jnl{Ap\&SS}}             
\def\aap{\aaref@jnl{A\&A}}                
\def\aapr{\aaref@jnl{A\&A~Rev.}}          
\def\aaps{\aaref@jnl{A\&AS}}              
\def\mnras{\aaref@jnl{Mon.~Not.~Roy.~Astron.~Soc.}}             
\def\prd{\aaref@jnl{Phys.~Rev.~D}}        
\def\prc{\aaref@jnl{Phys.~Rev.~C}}  
\def\prl{\aaref@jnl{Phys.~Rev.~Lett.}}    
\def\qjras{\aaref@jnl{QJRAS}}             
\def\skytel{\aaref@jnl{S\&T}}             
\def\ssr{\aaref@jnl{Space~Sci.~Rev.}}     
\def\zap{\aaref@jnl{ZAp}}                 
\def\nat{\aaref@jnl{Nature}}              
\def\aplett{\aaref@jnl{Astrophys.~Lett.}} 
\def\apspr{\aaref@jnl{Astrophys.~Space~Phys.~Res.}} 
\def\physrep{\aaref@jnl{Phys.~Rep.}}      
\def\physscr{\aaref@jnl{Phys.~Scr}}       
\def\commat{\aaref@jnl{Comm.~Math.~Phys.}}              
\def\science{\aaref@jnl{Science}}               
\def\cqg{\aaref@jnl{Classical Quant.~Grav.}}            
\def\jpcs{\aaref@jnl{JPCS}}                                     
\def\ijmpd{\aaref@jnl{Int.~J.~Mod.~Phys.~D}}                    
\def\grg{\aaref@jnl{Gen.~Relat.~Gravit.}}               
\def\rpp{\aaref@jnl{Rep.~Prog.~Phys.}}          
\def\npa{\aaref@jnl{Nucl.~Phys.~A}}        
\def\lrr{\aaref@jnl{Living Rev.~Rel.}}                   
\def\jcap{\aaref@jnl{J.~Cosmology Astropart.~Phys.}}    
\def\rmp{\aaref@jnl{Rev.~Mod.~Phys.}}   
\def\epjc{\aaref@jnl{Eur.~Phys.~J.~C}}

\allowdisplaybreaks[1]

\begin{document}

\color{black}       

\title{Constant sound speed and its thermodynamical interpretation in $f(Q)$ gravity}

\author{M. Koussour\orcidlink{0000-0002-4188-0572}}
\email{pr.mouhssine@gmail.com}
\affiliation{Quantum Physics and Magnetism Team, LPMC, Faculty of Science Ben
M'sik,\\
Casablanca Hassan II University,
Morocco.}

\author{Simran Arora\orcidlink{0000-0003-0326-8945}}
\email{dawrasimran27@gmail.com}
\affiliation{Department of Mathematics, Birla Institute of Technology and Science-Pilani,\\ Hyderabad Campus, Hyderabad-500078, India.}

\author{Dhruba Jyoti Gogoi\orcidlink{0000-0002-4776-8506}}
\email{moloydhruba@yahoo.in}
\affiliation{Department of Physics, Dibrugarh University,
Dibrugarh 786004, Assam, India.}

\author{M. Bennai\orcidlink{0000-0002-7364-5171}}
\email{mdbennai@yahoo.fr}
\affiliation{Quantum Physics and Magnetism Team, LPMC, Faculty of Science Ben
M'sik,\\
Casablanca Hassan II University,
Morocco.} 
\affiliation{Lab of High Energy Physics, Modeling and Simulations, Faculty of
Science,\\
University Mohammed V-Agdal, Rabat, Morocco.}

\author{P.K. Sahoo\orcidlink{0000-0003-2130-8832}}
\email{pksahoo@hyderabad.bits-pilani.ac.in}
\affiliation{Department of Mathematics, Birla Institute of Technology and
Science-Pilani,\\ Hyderabad Campus, Hyderabad-500078, India.}

\date{\today}
\begin{abstract}

On the basis of homogeneous and isotropic Friedmann-Lemaitre-Robertson-Walker (FLRW) geometry, solutions to the issues of cosmic acceleration and dark energy are being put forth within the context of $f\left( Q\right)$ gravity. We take into account a power law $f(Q)$ model using $f\left( Q\right) =\alpha Q^{n}$, where $\alpha $ and $n$ are free model parameters. In the current scenario, we may establish the energy density and pressure for our $f(Q)$ cosmic  model by applying the constant sound speed parameterizations, i.e., $\vartheta_{s}^{2}=\beta$, where a barotropic cosmic fluid is described in terms of $\beta$. The field equations are then derived, and their precise solutions are established. We obtain the constraints on the model parameters using the updated Hubble (Hz) data sets consisting of 31 data points, the recently published Pantheon samples (SNe) with 1048 points, and Baryon acoustic oscillations (BAO) data sets. We also examine the physical behaviour of the deceleration parameter, the equation of state (EoS) parameter, the statefinder diagnostic, and the Om diagnostic. We conclude that our $f\left( Q\right) $\ cosmic model predicts a transition in the universe from deceleration to acceleration. Further, to investigate the feasibility of the model, we discussed some of its thermodynamic aspects.

\end{abstract}

\maketitle

\section{Introduction}

\label{sec1}

General Relativity (GR) has successfully explained various aspects of the Universe, including gravitational waves, black holes, compact stars, etc. However, GR is not entirely free from issues and suffers significant problems in the UV, and infrared regime \cite{Sotiriou2010}. The theory and a number of observable findings, such as the accelerated expansion of the universe and galaxy rotation curves, are very different in the infrared spectrum. To deal with the infrared issues of GR, a simple but quite effective extension was suggested, which is known as the $\Lambda$CDM model. Although this model could explain the experimental deviations of GR, it is burdened with the presence of dark matter and dark energy. Dark matter and dark energy are unknown forms of matter and energy content of the Universe which have are not been directly detected until now. Moreover, the dark energy predicted by the $\Lambda$CDM model is static. The accelerated expansion of the Universe indicates the nature and properties of the unknown energy content, i.e. dark energy present in the Universe. It is mainly supported by observational studies like Type Ia supernovae \cite{Riess, Perlmutter}, baryon acoustic oscillations \cite{D.J., W.J.}, large-scale structure \cite{T.Koivisto, S.F.} and cosmic microwave background radiation \cite{R.R., Z.Y.}. Apart from the $\Lambda$CDM model, several models support the existence of such exotic matter and energy in the Universe and, as hypothesized by such models, around 
 $70\%$ of the Universe is filled with dark energy. It is worth to be mentioned that although the $\Lambda$CDM model was able to explain the observational results, it again comes with some drawbacks like the cosmic coincidence problem \cite{dalal2001}. It implies that the density of non-relativistic matter and dark energy are the same in the present Universe. Another issue with this model is the cosmological constant problem which shows a vast discrepancy between the astronomically observed values of cosmological constant $\Lambda$ \cite{Riess, Perlmutter} and theoretically predicted value of the quantum vacuum energy \cite{ weinberg/1989}. To overcome these issues, dynamical dark energy models like the Chaplygin gas model \cite{M.C., A.Y.}, k-essence  \cite{T.Chiba, C.Arm.}, quintessence \cite{Carroll, Fujii} etc. have been introduced. In these models, the energy-momentum part of the field
equations of GR is modified to explain the observational results.

There is another class of theories in which the geometrical part of the field equations of GR is modified. Such theories are termed modified theories of gravity (MTGs). Some of the promising MTGs are $f(R)$ gravity  \cite{Sotiriou2010}, $f(R,T)$ gravity, $f(R, L_m)$ gravity etc. The most simplified type of MTG is the $f(R)$ gravity, where the Ricci scalar in the action of the theory is replaced by a well-motivated function of the Ricci scalar \cite{Sotiriou2010}. Although higher-order terms in the gravity action have been previously included by Utiyama and De Witt \cite{Utiyama1962}, Buchdahl used the idea of $f(R)$ gravity for the first time in 1970 \cite{Buchdahl}.

Apart from such MTGs, there are another two approaches other than curvature representations \textit{viz.}, such as the teleparallel gravity and symmetric teleparallel gravity. In teleparallel gravity, the gravitational force is governed by the torsion $T$ \cite{Capozziello2011, Liu2012, Iorio2012, Wang2020, Nunes2016}. Einstein used the other approach, i.e. symmetric teleparallel gravity, and it was an attempt to unify field theories. Such theories account for vanishing curvature and torsion with non-vanishing non-metricity, which analyses how the length of a vector changes when paralely transported.

In this work, we shall use one of the promising MTGs known as $f(Q)$ gravity,
where the Lagrangian is a function of the non-metricity scalar $Q$ \cite{jimenez/2020}. One may note that $f(Q)$ gravity has obtained the attention of researchers in the last few years, \cite{khyllep/2021,mandal/2020,mandal/2020b,Koussour1,Koussour2,Koussour3} and a significant
number of studies have been done in this MTG to investigate different properties
of dark energy, including its evolution \cite{Yang2010, Bambda2011, Linder2010, Arora2022}. Here in this work, we consider homogeneous and isotropic FLRW geometry in the power law model of $f(Q)$ gravity and discuss the solutions to the issue of cosmic acceleration by constraining the model with observational data sets. For the completeness of the study, we also considered a black hole solution in the particular model and investigated its horizon thermodynamics in brief. Such an investigation will help us to comment on the viability of the model in terms of the thermodynamical aspects.

The paper is organized as follows. In section \ref{sec2}, we discussed the field equations and basics in $f(Q)$ gravity. We constructed the cosmological model with power law $f(Q)$ gravity model with constant sound speed parameterizations in section \ref{sec3}. The observational constraints on the model are obtained in section \ref{sec4}. We discussed the behaviour of cosmological parameters, such as the deceleration and equation of state parameters, and implemented the diagnostic methods like the statefinder diagnostic and $Om(z)$ diagnostics in section \ref{sec5}. In section \ref{sec6}, we briefly analyzed the thermodynamic parameters of a vacuum black hole solution in $f(Q)$ theory and studied its horizon thermodynamics and the first law. Finally, in section \ref{sec7}, we included a discussion and conclusion of our work.

\section{Some basics of $f(Q)$ gravity theory}

\label{sec2}

The action for $f(Q)$ gravity is written as,
\begin{equation}
S=\int \left( \frac{1}{2}f(Q)+L_{m}\right) \sqrt{-g}d^{4}x,  \label{1}
\end{equation}%
where $f(Q)$ is an arbitrary function related to the non-metricity scalar $%
Q $. In addition, $g=det(g_{\mu \nu })$ and $L_{m}$ denote the matter
Lagrangian. Furthermore, throughout this study, we will use units with the coupling constant $\kappa$ and the speed of light $c$ as 1. We further define the non-metricity scalar $Q$ as follows
\begin{equation}
Q\equiv -g^{\mu \nu }(L_{\,\,\,\alpha \mu }^{\beta }L_{\,\,\,\nu \beta
}^{\alpha }-L_{\,\,\,\alpha \beta }^{\beta }L_{\,\,\,\mu \nu }^{\alpha }),
\label{2}
\end{equation}%
where $L_{\,\,\,\alpha \gamma }^{\beta }$ is the disformation tensor, 
\begin{equation}
L_{\alpha \gamma }^{\beta }=-\frac{1}{2}g^{\beta \eta }(\nabla _{\gamma
}g_{\alpha \eta }+\nabla _{\alpha }g_{\eta \gamma }-\nabla _{\eta }g_{\alpha
\gamma }).  \label{3}
\end{equation}
The non-metricity tensor is given by 
\begin{equation}
Q_{\gamma \mu \nu }=\nabla _{\gamma }g_{\mu \nu },  \label{4}
\end{equation}
with the non-metricity traces as
\begin{equation}
Q_{\beta }=g^{\mu \nu }Q_{\beta \mu \nu }\qquad \widetilde{Q}_{\beta
}=g^{\mu \nu }Q_{\mu \beta \nu }.  \label{5}
\end{equation}

A superpotential or the non-metricity conjugate can also be defined as 
\begin{equation}
P_{\,\,\,\mu \nu }^{\beta }=-\frac{1}{2}L_{\,\,\,\mu \nu }^{\beta }+\frac{1}{%
4}(Q^{\beta }-\widetilde{Q}^{\beta })g_{\mu \nu }-\frac{1}{4}\delta _{(\mu
}^{\beta }Q_{\nu )}.  \label{6}
\end{equation}%
expressing the scalar of non-metricity as
\begin{equation}
Q=-Q_{\beta \mu \nu }P^{\beta \mu \nu }\,.  \label{7}
\end{equation}

Additionally, it is known that the energy-momentum tensor is defined by 
\begin{equation}
T_{\mu \nu }=-\frac{2}{\sqrt{-g}}\dfrac{\delta (\sqrt{-g}L_{m})}{\delta
g^{\mu \nu }}.  \label{8}
\end{equation}

We obtain the following field equations by equating the variation of action %
(\ref{1}) with respect to the metric to zero, 
\begin{widetext}
\begin{equation}
\frac{2}{\sqrt{-g}}\nabla _{\beta }\left( f_{Q}\sqrt{-g}P_{\,\,\,\,\mu \nu
}^{\beta }\right) +\frac{1}{2}fg_{\mu \nu }+f_{Q}(P_{\mu \beta \alpha
}Q_{\nu }^{\,\,\,\beta \alpha }-2Q_{\,\,\,\mu }^{\beta \alpha }P_{\beta
\alpha \nu })=-T_{\mu \nu },  \label{9}
\end{equation}%
\end{widetext}
where $f_{Q}=\dfrac{df}{dQ}$. One can obtain the following equation by varying the action in relation
to the connection.
\begin{equation}
\nabla _{\mu }\nabla _{\nu }(\sqrt{-g}f_{Q}P^{\mu \nu }{}_{\lambda })=0.
\label{2q}
\end{equation}

Recent CMB data show that our Universe is homogeneous and isotropic on a large scale, that is, on a scale more significant than that of galaxy clusters. For this reason, in the analysis we provide here, we take into consideration a flat FLRW background geometry in Cartesian coordinates with a metric,
\begin{equation}
ds^{2}=-dt^{2}+a^{2}(t)[dx^{2}+dy^{2}+dz^{2}],  \label{3a}
\end{equation}%
where $a(t)$ is the scale factor of the Universe. Additionally, the non-metricity scalar produced from the metric (\ref{3a}) is as follows:
\begin{equation}
Q=6H^{2},  \label{3b}
\end{equation}
where $H$ is the Hubble parameter, which measures the expansion rate of the Universe.

The perfect cosmic fluid, or cosmological fluid without taking into account viscosity effects, is the most frequently used energy-momentum tensor in cosmology. Hence, we have
\begin{equation}
T_{\mu \nu }=(\rho +p)u_{\mu }u_{\nu }+pg_{\mu \nu },  \label{3c}
\end{equation}
where $u^{\mu }=(1,0,0,0)$ denotes the four-velocity vector components that define the fluid, $\rho $ and $p$ denote, respectively, the cosmic energy density and isotropic pressure of the perfect cosmic fluid.

The $f(Q)$ gravity dynamics of the universe are described by modified
Friedmann equations, which are as follows. 
\begin{eqnarray}
\label{F1} 
3H^{2} &=&\frac{1}{2f_{Q}}\left( -\rho +\frac{f}{2}\right) ,  \\
\label{F2}
\dot{H}+3H^{2}+\frac{\dot{f}_{Q}}{f_{Q}}H &=& \frac{1}{2f_{Q}}\left( p+\frac{f}{2}\right) ,  
\end{eqnarray}
where an overhead dot points out the differentiation of the quantity with respect to the cosmic time $t$. It is important to note that if the function $f(Q)=-Q$ is assumed, the standard Friedmann equations of GR can be obtained.

We obtain the following evolution equation for $H$ by eliminating the term $3H^{2}$ thorugh the previous two equations.
\begin{equation}
\overset{.}{H}+\frac{\dot{f}_{Q}}{f_{Q}}H=\frac{1}{2f_{Q}}\left( p+\rho
\right) .  \label{HE}
\end{equation}

We can rewrite equations \eqref{F1} and \eqref{F2} and define the effective energy density and pressure as  
\begin{eqnarray*}
3 H^{2} &=& \rho + \rho_{Q},\\
3 H^{2}+ 2 \dot{H} &=& - \left(p + p_{Q}\right).
\end{eqnarray*}
Hence, we obatin
\begin{eqnarray*}
\rho_{Q} &=& 3 H^{2} \left( 1+ 2f_{Q} \right) -\frac{f}{2},\\
p_{Q} &=& - \left[ 2 \dot{H}\left(1+f_{Q} \right) -\frac{f}{2} + 3 H^{2}\left( 1+ 2 f_{Q} + 8 f_{QQ} \dot{H} \right) \right].
\end{eqnarray*}

Consequently, the matter conservation equation is obtained as given below.
\begin{equation}
\dot{\rho}+3H\left( \rho +p\right) =0.  \label{3f}
\end{equation}

According to the following equation of state, the cosmic fluid's normal
isotropic pressure and energy density are related by 
\begin{equation}
p=\omega \rho .  \label{EoS}
\end{equation}

Here, $\omega $ is the equation of state (EoS) parameter.

\section{Cosmological $f(Q)$ model with constant speed of sound}

\label{sec3}

For our investigation, we assume a specific power law model for the $f(Q)$ function, which is expressed as
\begin{equation}
f\left( Q\right) =\alpha Q^{n},  \label{f}
\end{equation}%
where $\alpha \neq 0$\ and $n$ are the model free parameters.

Using Eqs. (\ref{EoS}), (\ref{f}) and (\ref{HE}), we obtain a first-order differential equation for the Hubble parameter as 
\begin{equation}
\overset{.}{H}-\frac{H^{2\left( 1-n\right) }}{2\alpha n6^{n-1}\left(
2n-1\right) }\rho \left( \omega +1\right) =0.  \label{ED}
\end{equation}

Since the isotropic pressure and energy density of a barotropic fluid are related, the EoS can be stated implicitly as

\begin{equation}
G\left( \rho ,p\right) =0.  \label{G}
\end{equation}

Thus, using Eq. (\ref{EoS}), one can write Eq. (\ref{G}) as $F(\rho ,\omega )=0$ and $G(\rho ,\omega
)=F(\rho ,p)$. We can think of the energy density $\rho =\rho \left( \omega \right) $ and isotropic pressure $p=p\left( \omega \right) =\omega \rho \left( \omega \right) $ as functions of $\omega $. Moreover, the inversion of $F(\rho ,\omega )=0$ suggests that other solutions to the equations $p\left( \omega \right) $ and $\rho \left( \omega \right) $ can be derived. There may be numerous values of $\rho \left( \omega \right)$, particularly for specific values of $\omega $. One of the strictest tests
to determine whether a cosmological model is valid is the speed of sound $\vartheta _{s}^{2}$. If the speed of sound $\vartheta _{s}^{2}$ is lower than the speed of light $c$, a model is considered to be physically plausible. The relation $0\leq \vartheta _{s}^{2}\leq c$ specifies the stability prerequisite for the cosmological models. In this study, we have assumed that the speed of light is $c=1$. Thus, if condition $0\leq \vartheta _{s}^{2}\leq 1$ is met, the model is physically plausible. 
These constraints make this kind of modeling more appropriate, and certain
models with variable sound speed have been described in the literature \cite{sp1, sp2, sp3, sp4, sp5}. The squared speed of sound ($\vartheta _{s}^{2}$) in barotropic cosmic fluid can be described as
\begin{equation}
\vartheta _{s}^{2}=\frac{dp}{d\rho }.
\end{equation}

Differentiating Eq. (\ref{G}) gives
\begin{equation}
\frac{\partial G}{\partial \rho }d\rho +\frac{\partial G}{\partial p}dp=0,
\label{G1}
\end{equation}
which brings about
\begin{equation}
\vartheta _{s}^{2}=-\frac{\frac{dG}{d\rho }}{\frac{dG}{dp}}.  \label{G2}
\end{equation}

Using Eqs. (\ref{EoS}), (\ref{G1}), and Eq. (\ref{G2}), we have 
\begin{equation}
\frac{d\rho }{\rho }=\frac{d\omega }{\vartheta _{s}^{2}-\omega }.  \label{G3}
\end{equation}

Combining Eqs. (\ref{G3}) and (\ref{3f}), one can obtain
\begin{equation}
\frac{d\omega }{\left( \vartheta _{s}^{2}-\omega \right) \left( 1+\omega
\right) }=3\frac{dz}{1+z},  \label{G4}
\end{equation}%
where we used $\frac{dz}{dt}=-\left( 1+z\right) H$. Since, we know that $\rho $ and $p$ are functions of $\omega $, the sound speed, $\vartheta_{s}^{2}$ can also be thought of as a function of $\omega $, i.e., $\vartheta_{s}^{2}=\vartheta _{s}^{2}\left( \omega \right) $
\begin{equation}
\vartheta _{s}^{2}=\frac{dp}{d\rho }=\frac{\frac{dp}{d\omega }}{\frac{d\rho 
}{d\omega }}.
\end{equation}

Thus, Eq. (\ref{G4}) governs the dynamics of the EoS parameter $\omega $.
Here, we consider a constant sound speed parameterizations \cite{sp6, sp7, sp8},%
\begin{equation}
\vartheta _{s}^{2}=\beta ,
\end{equation}%
where $\beta $\ is a constant. Integration of Eq. (\ref{G4}) generate the
parameter of EoS as%
\begin{equation}
\omega \left( z\right) =\frac{\beta \frac{1+\omega _{0}}{\beta -\omega _{0}}%
\left( 1+z\right) ^{3\left( 1+\beta \right) }-1}{\frac{1+\omega _{0}}{\beta
-\omega _{0}}\left( 1+z\right) ^{3\left( 1+\beta \right) }+1}.
\end{equation}

By integrating Eq. (\ref{G3}), we obtain the following relation
\begin{eqnarray}
 \label{G5}
\rho &=&\rho _{0}\frac{\beta -\omega _{0}}{\beta -\omega },\\ 
 \label{G6}
p &=& \beta \rho -\rho _{0}\left( \beta -\omega _{0}\right),
\end{eqnarray}
where $\omega \left( 0\right) =\omega _{0}$. Eqs. (\ref{G5}) and (\ref{G6}) can further used to obtain energy density and isotropic pressure as follows.
\begin{equation}
\rho \left( z\right) =\rho _{0}\frac{\beta -\omega _{0}}{1+\beta }\left( 
\frac{1+\omega _{0}}{\beta -\omega _{0}}\left( 1+z\right) ^{3\left( 1+\beta
\right) }+1\right) ,  \label{G7}
\end{equation}%
\begin{equation}
p\left( z\right) =\rho _{0}\frac{\beta -\omega _{0}}{1+\beta }\left( \beta 
\frac{1+\omega _{0}}{\beta -\omega _{0}}\left( 1+z\right) ^{3\left( 1+\beta
\right) }-1\right) .
\end{equation}

Further, we can define the relation for $t$ and $z$ using the formula $%
a=a_{0}\left( 1+z\right) ^{-1}$, as shown below,%
\begin{equation}
\frac{d}{dt}=\frac{dz}{dt}\frac{d}{dz}=-\left( 1+z\right) H\left( z\right) 
\frac{d}{dz}.
\end{equation}

Setting the present value of scale factor to $a_{0}=a(0)=1$ as a standard.
The Hubble parameter can be expressed mathematically as,%
\begin{equation}
\overset{.}{H}=-\left( 1+z\right) H\left( z\right) \frac{dH}{dz}.
\end{equation}

Now, by resolving Eq. (\ref{ED}), in terms of redshift, we found the
following expression for the Hubble parameter:
\begin{equation}
H(z)=\left[ H_{0}^{2n}+\frac{2(6^{-n})(1+\omega _{0})\rho
_{0}(1-(1+z)^{3+3\beta })}{(2n-1)\alpha (1+\beta )}\right] ^{\frac{1}{2n}}
\label{H}
\end{equation}
where $H_{0}$ is the present value of the Hubble parameter.

\section{Observational constraints}

\label{sec4}

This section presents the cosmological constraints of the considered model. The
statistical method we use helps us to constrain the parameters such as $%
\alpha$, $\beta$, $n$, $\omega_{0}$, $H_{0}$, and $\rho_{0}$. We chose the
Markov Chain Monte Carlo (MCMC) with the conventional Bayesian approach. The following data sets are used:
\begin{itemize}
    \item \textbf{Hubble data:} We use a standard collection of 31 measurements obtained from the differential age method (DA) \cite{Moresco,Ratra}. The DA method is employed to calculate the rate of expansion at redshift $z$. The following formula is used to determine chi-square ($\chi^2$). 
    \begin{equation}
        \chi^2_{Hz}= \sum_{j=1}^{31} \frac{\left[H(z_{j})-H_{obs}(z_{j}, p_{s})\right]^2}{\sigma(z_{j})^2}. 
    \end{equation}
 Here, $H_{obs}$ represents the observational value, $p_{S}$ is the parameter space. $\sigma^{2}$ is the observed error. 
    \item \textbf{SNe Ia data:} The supernovae (SNe Ia) observation is crucial to understand how the universe is expanding. Significantly, the SNe Ia data is recorded from the Panoramic Survey Telescope and Rapid Response system (Pan-STARSS1), Sloan Digital Sky Survey (SDSS), Supernova Legacy Survey (SNLS), and Hubble Space Telescope (HST) survey \cite{Scolnic}. We use the Pantheon sample consisting of 1048 points of distance modulus $\mu_{j}$ in the range $0.01<z_{j}<2.26$ at different redshift. We perform the analysis using the expressions 
    \begin{eqnarray}
    \mu^{th}(z_{j}) &=& 25 +  5 log_{10}\left[\frac{d_{l}(z)}{1 Mpc}\right], \\
    d_{l}(z) &=& c(1+z) \int_{0}^{z} \frac{dy}{H(y,p_s)},\\
    \chi^{2}_{SN} &=& \sum_{j,i=1}^{1048} \Delta \mu_{j} (C_{SN}^{-1})_{ji} \Delta \mu_{i}  .
    \end{eqnarray}
Here, $\Delta \mu_{j}=u_{th}(z_{j},p_{s})-\mu_{obs} $, $p_{s}$ is the parameter space, $C_{SN}$ is the covariance matrix.     
    
\item \textbf{BAO data:} We consider the sample from $SDSS$, $6dFGS$, $Wiggle$ Z surveys at various redshifts. The following cosmology to establish BAO constraints $\left(\frac{d_A(z)}{D_{v}(z)}\right)$ are as follows:
\begin{eqnarray}
d_{A}(z) &=& c\int_{0}^{z}\frac{dx }{H(x,p_{s})},\\
D_{v}(z) &=& \left[\frac{d_{A}^2(z) c z}{H(z)} \right]^\frac{1}{3},\\
\chi^{2}_{BAO} &=& Y^{T} C_{BAO}^{-1} Y.
\end{eqnarray}
where $Y$ depends on the survey considered and $C_{BAO}$ is the covariance matrix \cite{Giostri}. 
 
 \item {\textbf{Results:}}  The constraints on the model parameters for the joint ($Hz+SNe+BAO$) are obtained using $\chi^2= \chi^{2}_{Hz}+ \chi^{2}_{BAO}+ \chi^{2}_{SNe} $. The outcomes and results are shown in Table \ref{tab}. Additionally, figures \ref{fig1} and \ref{fig2} illustrate the likelihood contours for $Hz$, $SNe$ and $Joint$ analysis. One can observe that the observations from $Hz$ and $SNe$ are more consistent than the joint $Hz+SNe+BAO$ data-sets. It is worth mentioning that the values of parameter $H_{0}$ align with the observations \cite{Aghanim}. 
\end{itemize}

\begin{widetext}

\begin{figure}[h]
\centerline{\includegraphics[scale=0.5]{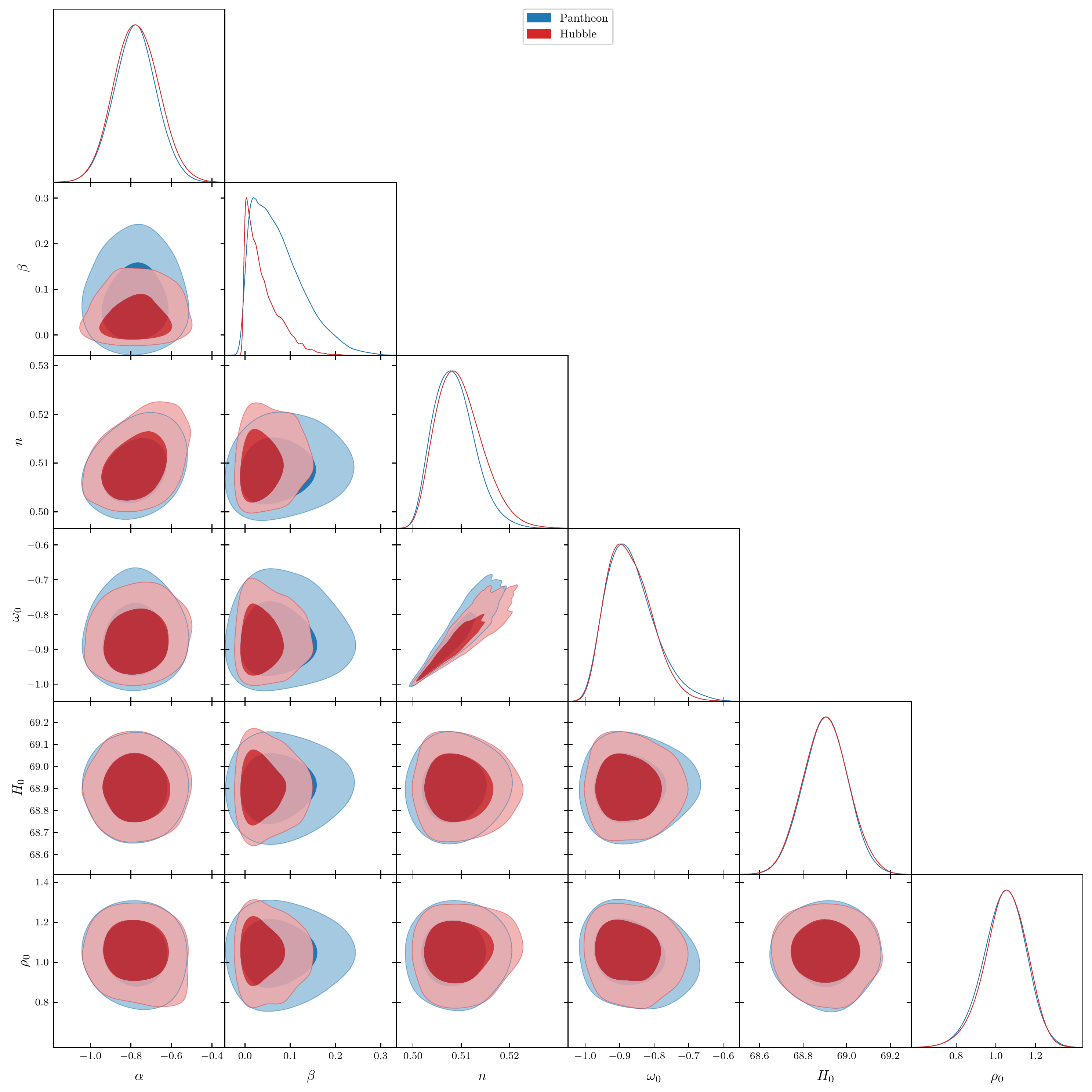}}
\caption{The $1-\sigma$ and $2-\sigma$ confidence regions of parameter space using Hubble  and Pantheon samples.}
\label{fig1}
\end{figure}

\begin{figure}[h]
\centerline{\includegraphics[scale=0.6]{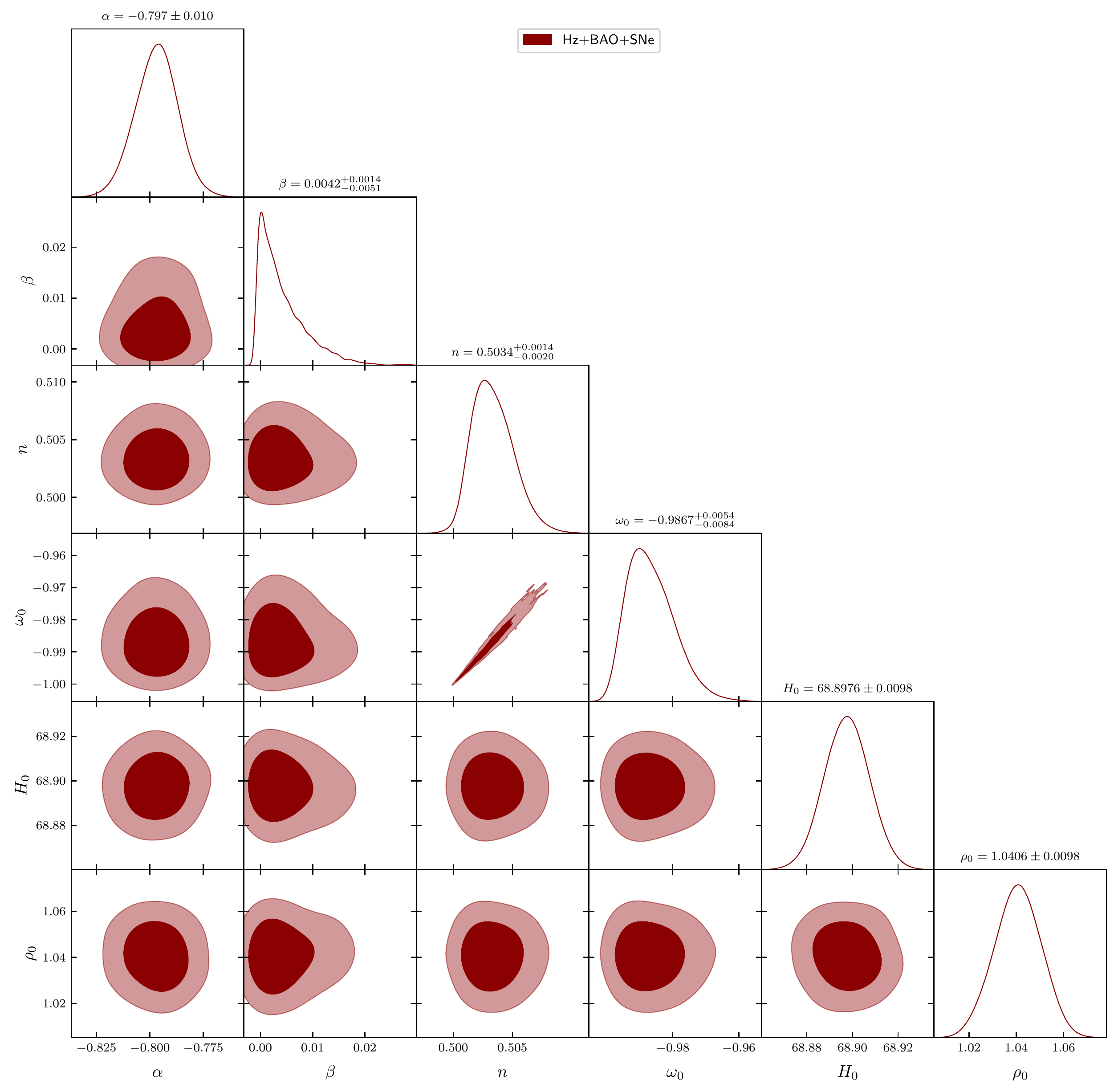}}
\caption{The $1-\sigma$ and $2-\sigma$ confidence regions of parameter space using $Hz+SNe+BAO$ sample.}
\label{fig2}
\end{figure}

\begin{table*}[!htbp]
\begin{center}
\begin{tabular}{l c c c c c c}
\hline 
data-sets              & $\alpha$ & $\beta$ & $n$ & $\omega_0$ & $H_{0}$ & $\rho_{0}$ \\
\hline
$Hubble (Hz)$ & $-0.77^{+0.11}_{-0.11}$  & $0.039^{+0.011}_{-0.043}$  & $0.5097^{+0.0037}_{-0.0056}$ & $-0.872^{+0.052}_{-0.075}$ & $68.90^{+0.10}_{-0.10}$ & $1.05^{+0.11}_{-0.094}$\\
$Pantheon (SNe)$   & $-0.78^{+0.10}_{-0.10}$  & $0.075^{+0.027}_{-0.073}$  & $0.5085^{+0.0034}_{-0.0048}$ & $-0.868^{+0.051}_{-0.083}$ & $68.90^{+0.10}_{-0.10}$ & $1.04^{+0.11}_{-0.099}$\\
$Hz+SNe+BAO$   & $-0.797^{+0.010}_{-0.010}$  & $0.0042^{+0.0014}_{-0.0051}$  & $0.5034^{+0.0014}_{-0.0020}$ & $-0.9867^{+0.0054}_{-0.0084}$ & $68.8976^{+0.0098}_{-0.0098}$ & $1.0406^{+0.0098}_{-0.0098}$\\

\hline
\end{tabular}
\caption{Best-fit values of model parameters}
\label{tab}
\end{center}
\end{table*}

\end{widetext}

\section{Cosmological parameters}

\label{sec5}

In modern cosmology, studying cosmological parameters has attracted much interest in understanding the expansion dynamics of the Universe better. This part will explore the cosmological parameters for the earlier built model, including the deceleration parameter, EoS parameter, statefinder diagnostics, and Om diagnostic.

\subsection{Deceleration parameter}

One of the crucial elements needed to explain the behavior of the Universe is the deceleration parameter ($q$). The sign of the deceleration parameter, which can be negative or positive, determines whether the Universe is accelerating or decelerating. The definition of the deceleration parameter
is
\begin{equation}
q=-1-\frac{\overset{.}{H}}{H^{2}}.
\end{equation}

According to our cosmological $f\left( Q\right) $\ model, the deceleration
parameter is obatined as  
\begin{widetext}
\begin{equation}
q\left( z\right) =\frac{3(\beta +1)\text{$\rho $}_{0}(\text{$\omega $}%
_{0}+1)(z+1)^{3\beta +3}}{n\left( 2\text{$\rho $}_{0}(\text{$\omega $}%
_{0}+1)\left( z^{3}(z+1)^{3\beta }+3z^{2}(z+1)^{3\beta }+3z(z+1)^{3\beta
}+(z+1)^{3\beta }-1\right) -\alpha (\beta +1)6^{n}(2n-1)H_{0}^{2n}\right) }-1
\label{qz}
\end{equation}
\end{widetext}

\begin{figure}[h]
\centerline{\includegraphics[scale=0.55]{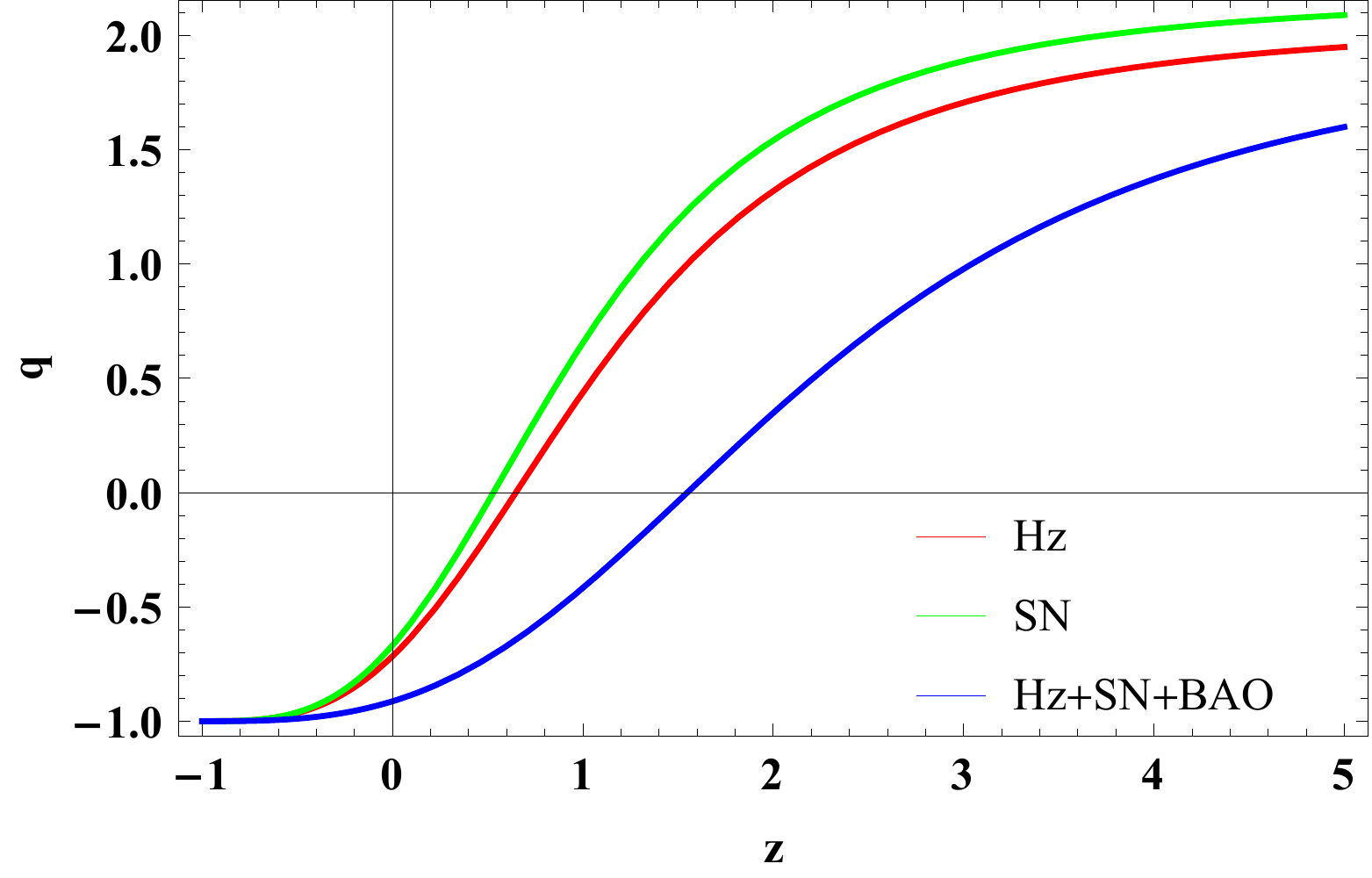}}
\caption{The graphical behavior of the deceleration parameter with the
constraint values from the $Hz$, $SNe$, and $Hz+SNe+BAO$ data-sets.}
\label{fig_q}
\end{figure}

According to the values of model parameters imposed by the $Hz$, $SNe$, and
$Hz+SNe+BAO$ data-sets, Fig. \ref{fig_q} illustrates the behavior of the
deceleration parameter $q$ versus redshift $z$. It shows that our cosmic $
f(Q)$ model is capable of producing both the early deceleration expansion ($
q>0$) and the late-time cosmic acceleration ($q<0$). Also, for the data-sets
from $Hz$, $SNe$, and $Hz+SNe+BAO$, the deceleration parameter currently has values of $q_{0}=-0.71^{+0.07}_{-0.04}$, $q_{0}=-0.66^{+0.06}_{-0.09}$, and $q_{0}=-0.91^{+0.00005}_{-0.008}$, respectively \cite{Capo,Camarena}.

\subsection{Equation of State parameter}

As seen above, the relation between energy density $\rho $ and isotropic
pressure $p$ is called the EoS parameter denoted by $\omega $. The EoS
parameter is employed to characterize the accelerated and decelerated
expansion of the Universe, and it divides different epochs into three
categories: The radiation-dominated phase is shown by the model when $\omega
=\frac{1}{3}$, the matter-dominated phase by $\omega =0$, and the stiff
fluid phase by $\omega =1$. In the current stage of accelerated evolution, $%
-1<\omega \leq -\frac{1}{3}$, indicates the quintessence phase, $\omega =-1$
indicates the cosmological constant, or $\Lambda$CDM model, and $\omega <-1$%
\ indicates the phantom era.

\begin{figure}[h]
\centerline{\includegraphics[scale=0.55]{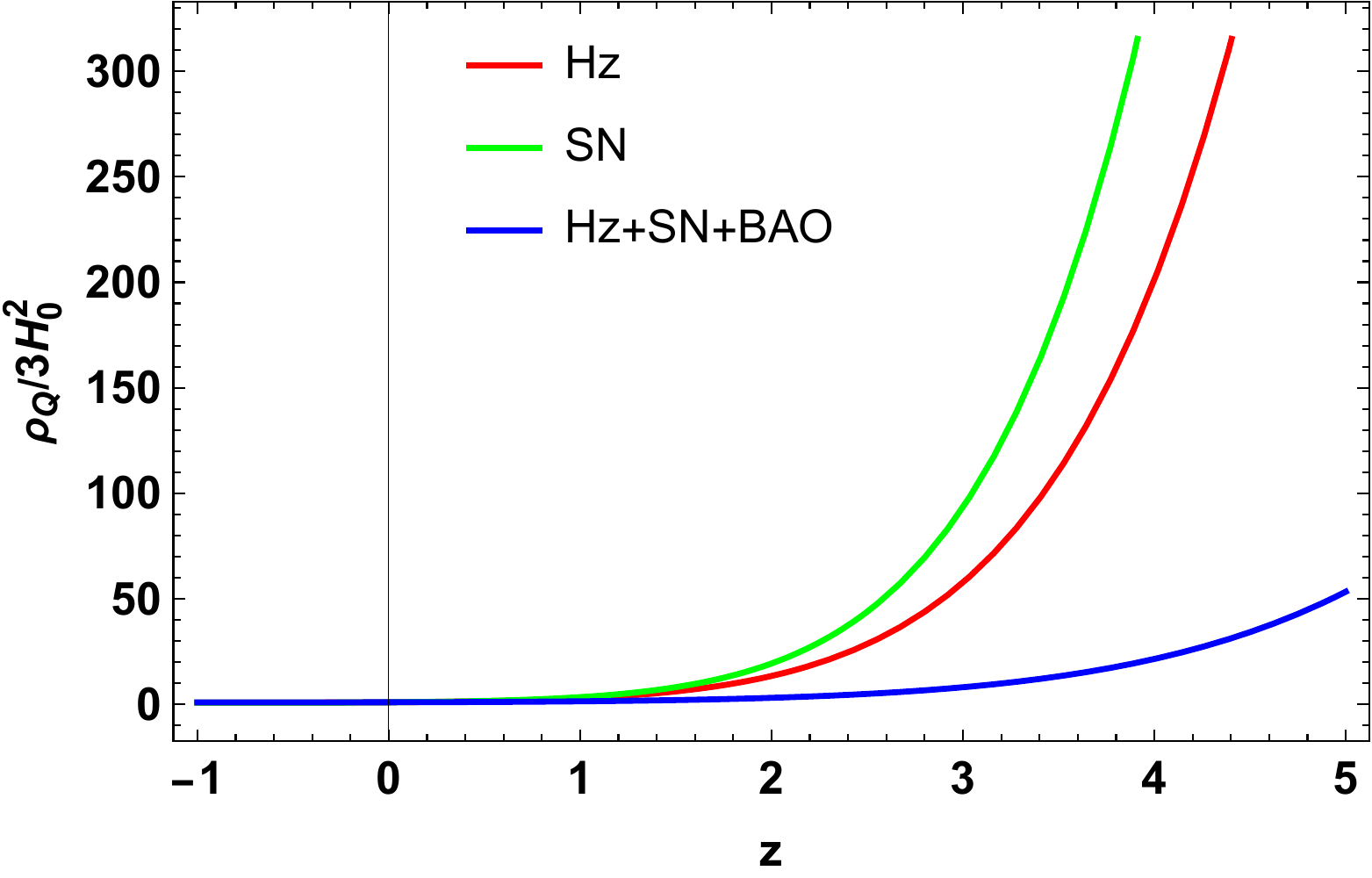}}
\caption{The graphical behavior of the density parameter for the non-metricity component with the constraint
values from the $Hz$, $SNe$, and $Hz+SNe+BAO$ data-sets.}
\label{fig_rhoQ}
\end{figure}

\begin{figure}[h]
\centerline{\includegraphics[scale=0.55]{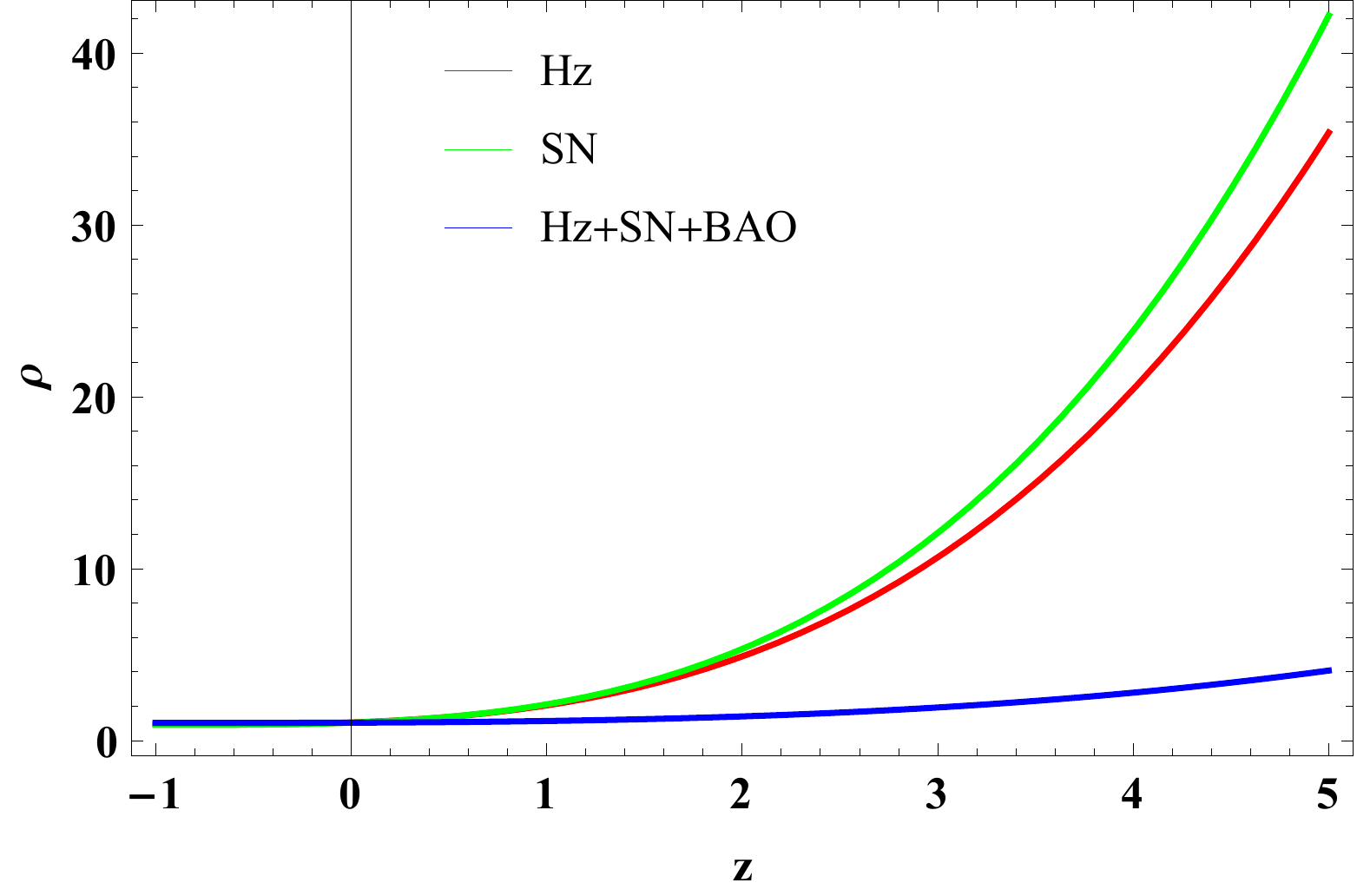}}
\caption{The graphical behavior of the energy density with the constraint
values from the $Hz$, $SNe$, and $Hz+SNe+BAO$ data-sets.}
\label{fig_rho}
\end{figure}

\begin{figure}[h]
\centerline{\includegraphics[scale=0.55]{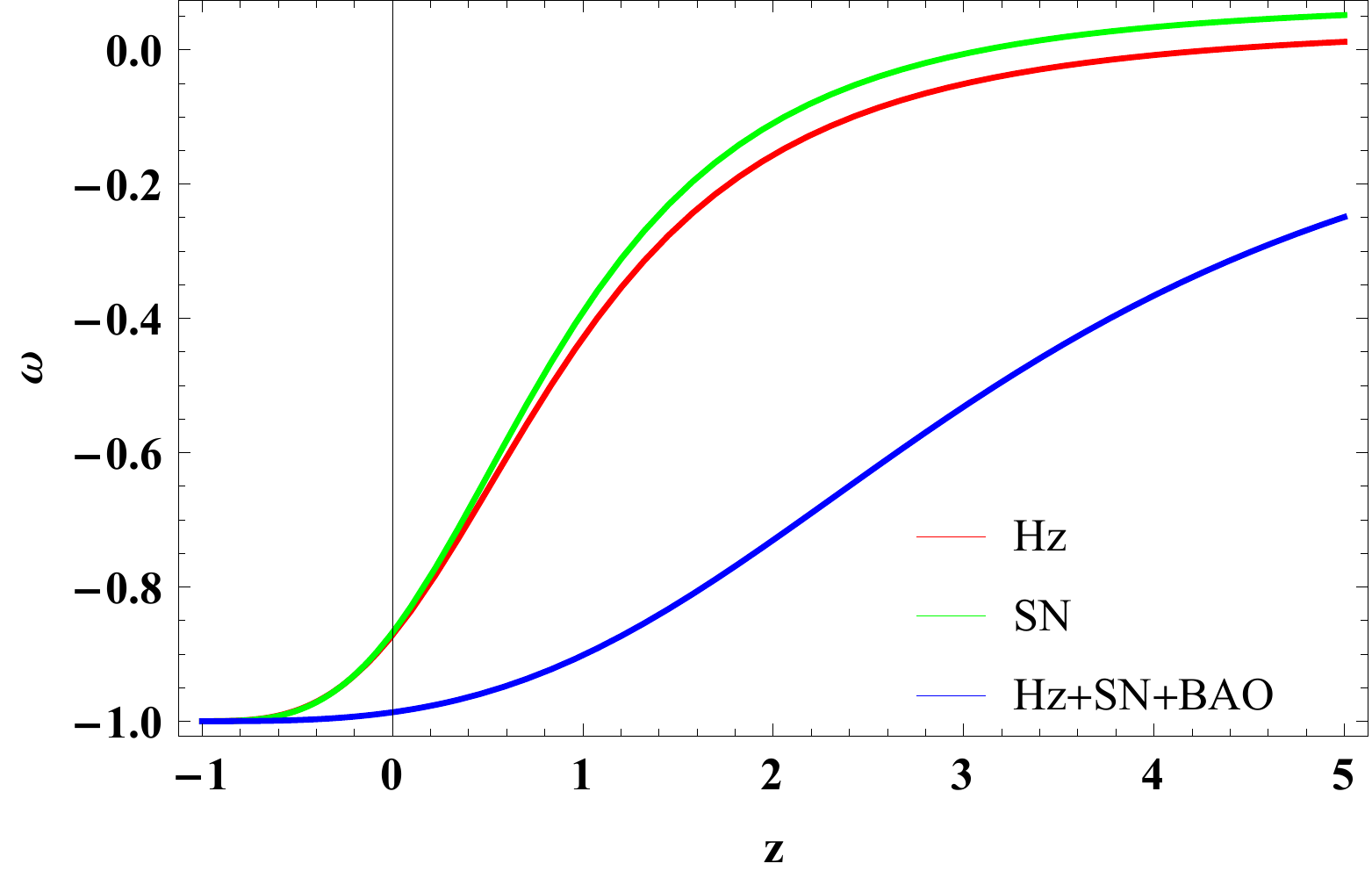}}
\caption{The graphical behavior of the EoS parameter with the constraint
values from the $Hz$, $SNe$, and $Hz+SNe+BAO$ data-sets.}
\label{fig_omega}
\end{figure}

Figs. \ref{fig_rhoQ} and \ref{fig_rho} depict the behavior of both density parameter for the non-metricity component and energy density of the universe for the
parameter values constrained by the $Hz$, $SNe$, and $Hz+SNe+BAO$ data-sets,
respectively. It can be shown that the two densities behave positively
with redshift $z$ for all data-sets. Further, the EoS parameter as seen in
Fig. \ref{fig_omega} suggests that our cosmic $f(Q)$ model with a constant
speed of sound behaves in a similar way to quintessence dark energy for
larger values of $z$ and approaches the $\Lambda$CDM point for lower values
of $z$. The current values of the EoS parameter are $\omega
_{0}=-0.872_{-0.075}^{+0.052}$, $\omega _{0}=-0.868_{-0.083}^{+0.051}$, and $
\omega _{0}=-0.9867_{-0.0084}^{+0.0054}$, respectively, for the $Hz$, $SNe$, and $Hz+SNe+BAO$ data-sets \cite{Santos}.

\subsection{Statefinder diagnostics}

Sahni et al. developed the statefinder cosmological diagnostic pair $\left\{
r,s\right\} $ in \cite{Sahni}. Similar to the geometrical parameters $H(z)$
and $q(z)$ discussed in previous sections, the parameters $r$ and $s$ are
dimensionless and are created from the scale factor of the Universe $a(t)$
and its temporal derivatives. The statefinder makes it easier to distinguish
and contrast various dark energy scenarios. As shown below, there are certain fixed points,
in the $s-r$ plane and $q-r$ plane in the cosmological constant model ($\Lambda $CDM) and the standard cold dark model (SCDM). Any obtained model can be checked to these standard models to
determine how closely it conforms to or differs from them. Following is a
definition of these parameters:
\begin{eqnarray}
r &=&\frac{\overset{...}{a}}{aH^{3}},\\
s &=&\frac{\left( r-1\right) }{3\left( q-\frac{1}{2}\right) }.
\end{eqnarray}

The parameter $r$ can be rewritten as
\begin{equation}
r=2q^{2}+q-\frac{\overset{.}{q}}{H}.
\end{equation}

The statefinder pair $\left\{ r,s\right\} $ represents the following dark
energy models for various values:

\begin{itemize}
\item $\Lambda $CDM model is equivalent to ($r=1,s=0$),

\item Holographic dark energy model is equivalent to ($r=1,s=\frac{2}{3}$),

\item Chaplygin gas model is equivalent to ($r>1,s<0$),

\item Quintessence model is equivalent to ($r<1,s>0$),
\end{itemize}

\begin{figure}[h]
\centerline{\includegraphics[scale=0.55]{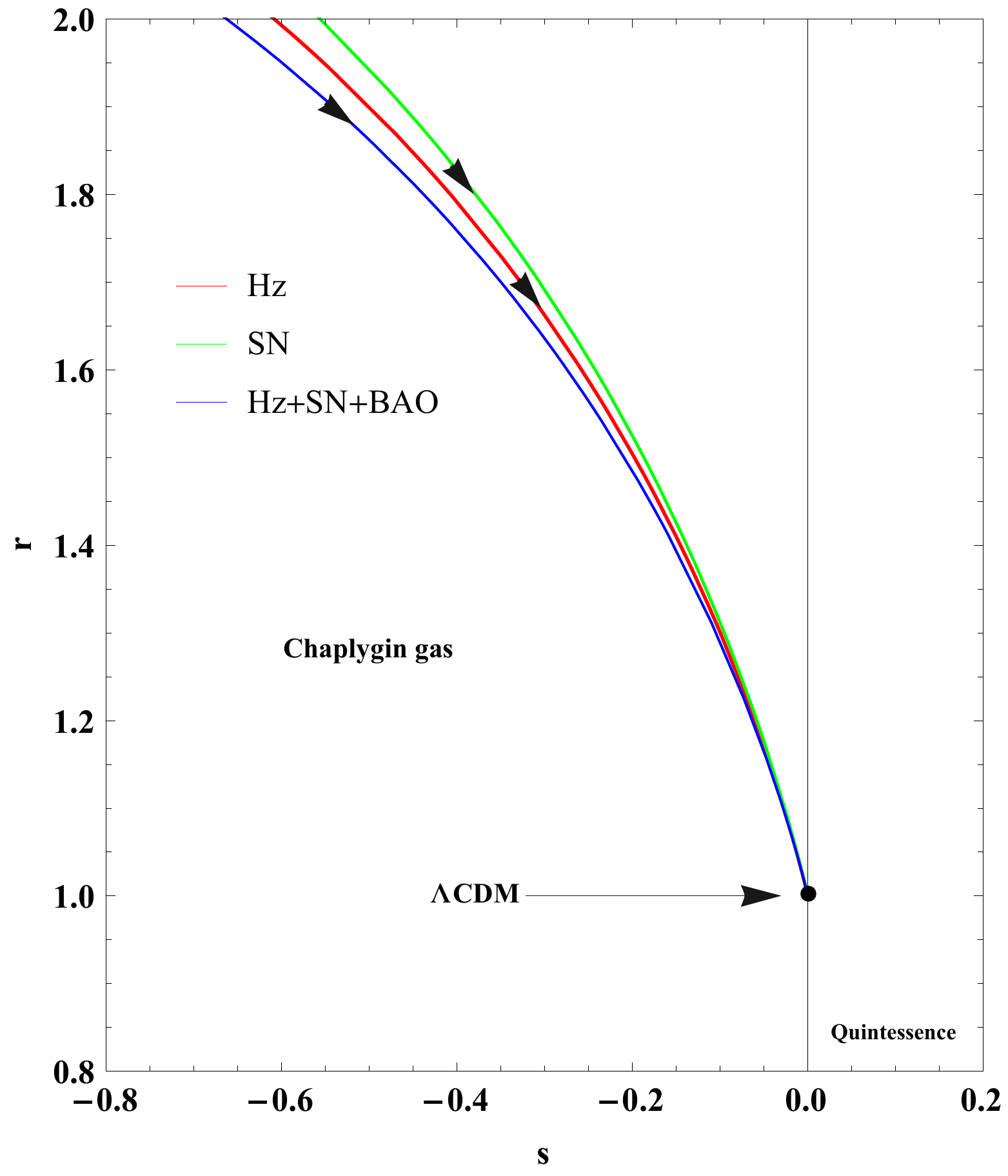}}
\caption{The graphical behavior of the $r-s$ plan with the constraint values
from the $Hz$, $SNe$, and $Hz+SNe+BAO$ data-sets.}
\label{fig_rs}
\end{figure}

\begin{figure}[h]
\centerline{\includegraphics[scale=0.55]{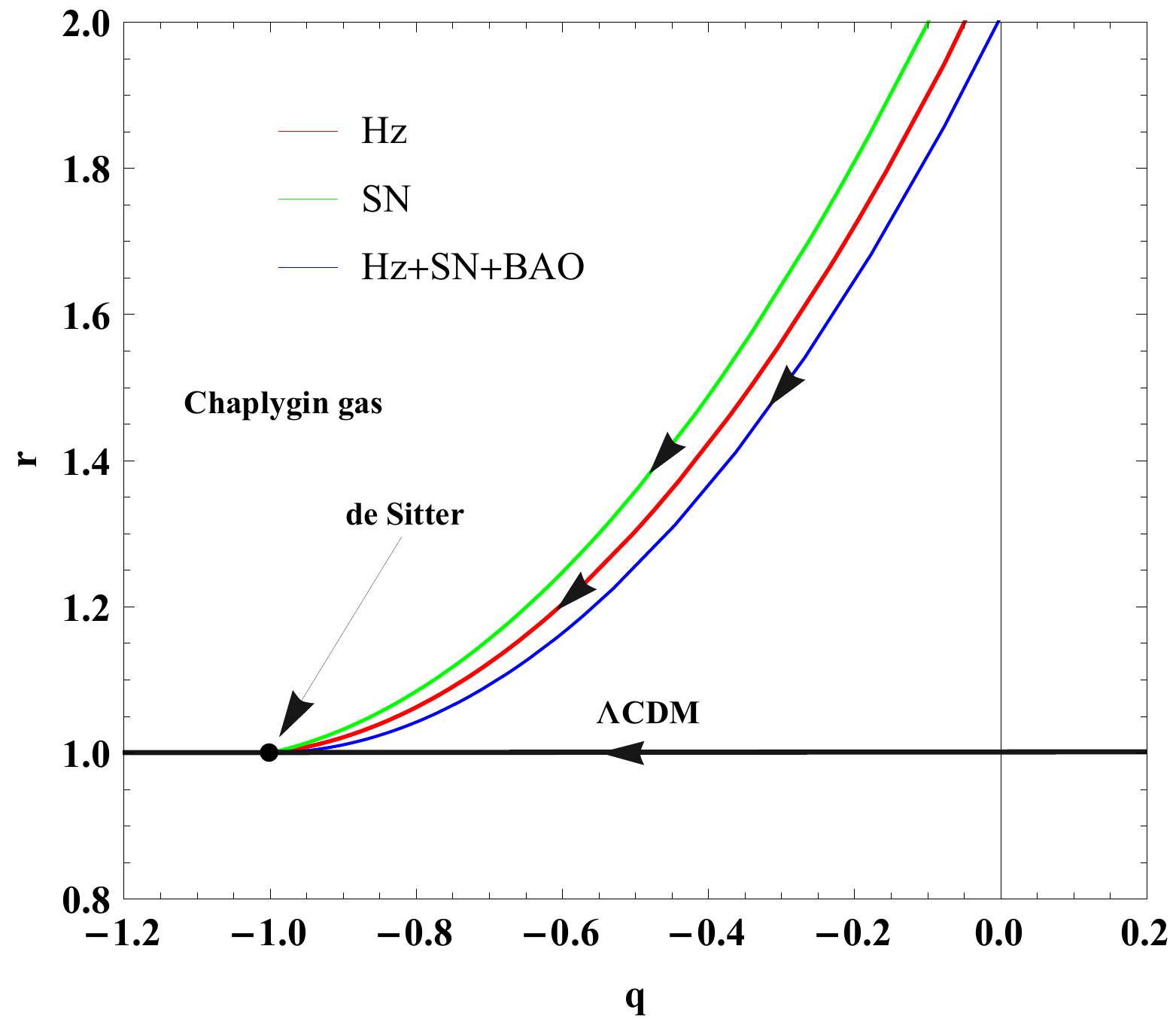}}
\caption{The graphical behavior of the $r-q$ plan with the constraint values
from the $Hz$, $SNe$, and $Hz+SNe+BAO$ data-sets.}
\label{fig_rq}
\end{figure}

The $s-r$ and $q-r$ graphs for our cosmic $f(Q)$ model are presented in
Figs. \ref{fig_rs} and \ref{fig_rq} using the values of the parameters
imposed by the $Hz$, $SNe$, and $Hz+SNe+BAO$ data-sets. Fig. \ref{fig_rs} shows that
the trajectory initially departs from the $\Lambda $CDM model before
eventually converging to it. Also, the Chaplygin gas zone (which is
symbolized by $r>1,s<0$) perfectly accounts for the trajectory's evolution.
According to Fig. \ref{fig_rq}, our model begins with the Chaplygin gas and
moves on to the de-Sitter point ($q=-1,r=1$) at the end. Therefore, the
statefinder diagnostic effectively demonstrates how the provided model
differs from other DE models.

\subsection{Om diagnostic}

We will now describe the $Om$ diagnostic, known as $Om(z)$. The typical 
$\Lambda $CDM model is distinguished from numerous dark energy models using $
Om(z)$. Due to the Hubble parameter dependence on a single temporal
derivative of $a(t)$ function, only first-order derivatives are employed in
the study of $Om$ diagnostic. The definition of $Om(z)$ for a flat Universe
is \cite{Sahni1, Zunckel},
\begin{equation}
Om\left( z\right) =\frac{\left( \frac{H\left( z\right) }{H_{0}}\right) ^{2}-1%
}{\left( 1+z\right) ^{3}-1}.  \label{Om}
\end{equation}

As a result, the $\Lambda $CDM model, phantom, and quintessence cosmological
models all have different values for $Om(z)$. We can categorize the behavior
of dark energy as quintessence type ($\omega >-1$), which corresponds to its
negative slope, phantom type ($\omega <-1$), which corresponds to its
positive slope, and $\Lambda $CDM type ($\omega =-1$), which corresponds to
zero slopes.

\begin{figure}[h]
\centerline{\includegraphics[scale=0.55]{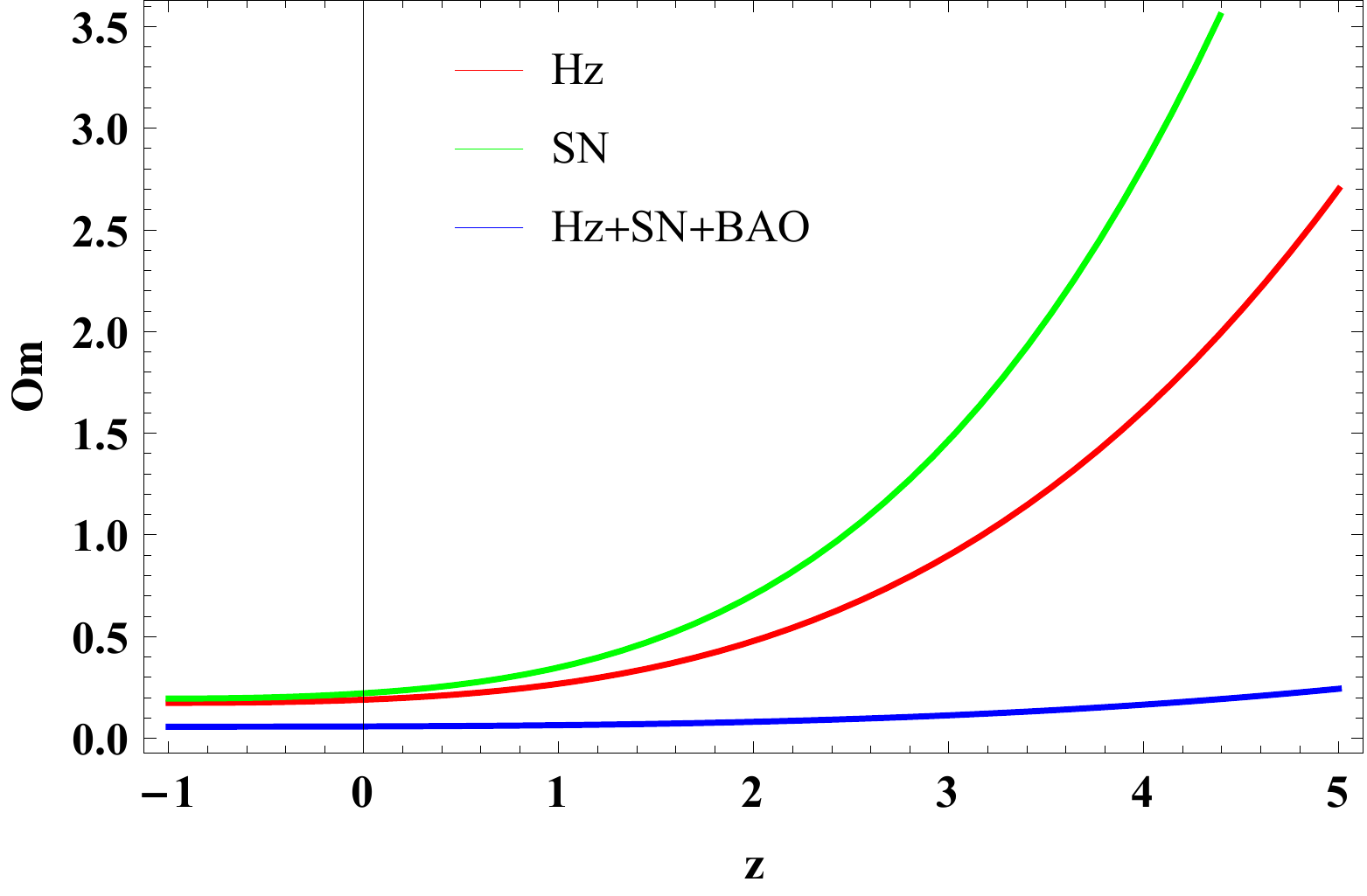}}
\caption{The graphical behavior of the Om diagnostic with the constraint
values from the $Hz$, $SNe$ and $Hz+SNe+BAO$ data-sets.}
\label{fig_om}
\end{figure}

Fig. \ref{fig_om} shows the positive slope throughout the entire range of
the $Om(z)$ diagnostic parameter for the constrained values of the model
parameters by the $Hz$, $SNe$, and $Hz+SNe+BAO$ data-sets. Our cosmic $f(Q)$ model
thus exhibits phantom-type behavior, according to the $Om(z)$ diagnostic
test.

\section{Thermodynamics aspects of the model}

\label{sec6}

We consider the following ansatz for a spherically symmetric static black hole in $f(Q)$ gravity,
\begin{equation}
    ds^2 = - h(r) dt^2 + 1/g(r) dr^2 + r^2 (d \theta^2 + \sin^2 \theta d\phi^2).
\end{equation}

For this case, we have the non-metricity scalar given by,
\begin{equation}
    Q = \frac{(g(r)-1) \left(g(r) h'(r)-h(r) g'(r)\right)}{r h(r) g(r)}.
\end{equation}

One may note that the non-metricity scalar shows that one can't simply choose $h(r) = g(r)$ as the case for the Schwarzschild black hole because this makes the non-metricity scalar vanish, i.e., $Q=0$ \cite{Lin21}. Hence, for this analysis, we shall pick $h(r) \ne g(r)$ to ensure that $Q$ survives. Now we follow the Ref. \cite{BHfQ}, where a black hole solution of the following form has been obtained for the power law model:
\begin{eqnarray}
h(r) &=& \left(\frac{r}{r_T}\right)^\beta \left[ 1 - \left( \frac{r_s}{r} \right)^{-\gamma} \right],\\
g(r) &=& \frac{1}{C} \left[ 1 - \left( \frac{r_s}{r} \right)^{-\gamma} \right].
\end{eqnarray}

Here $r_T$, $\beta$, $\gamma$ and $C$ are constants associated with the solution and $r_s$ stands for the horizon radius. One may note that this black hole solution is not physically viable \cite{BHfQ} and suffers from several issues. Till now, no other black hole solutions have been obtained in the power law model of $f(Q)$ gravity which is physically viable. Hence, we pick this solution for a brief qualitative analysis to see how this model and non-metricity may affect the horizon thermodynamics and the related parameters. Detailed analysis, as well as physical viability of the black hole solutions, are kept as future scope of the study.
From previous studies, it is evident that the thermodynamics is approximately identical for both event and cosmological horizon \cite{Bousso}. So we can use the above metric functions of the black hole space-time to study thermodynamics for the power law model of $f(Q)$ gravity. Several studies deal with a flat universe to check the same in different frameworks \cite{Bousso, FR,UH}.

We can study the horizon thermodynamics of the theory by following Refs. \cite{newref1, newref2}. However, here we have considered the power law model of $f(Q)$ gravity framework, for which no physically viable black hole solution still has been obtained \cite{BHfQ}. A rigorous study of black holes in this new theory remains in the literature. Therefore, in this study, we discuss a few properties in brief only. For this purpose, we consider the black hole solution mentioned above. One may note that the above solution of the black hole for power law $f(Q)$ gravity does not have a Schwarzschild limit, and at an infinite distance away from the black hole horizon, it can't provide a Minkowski space-time \cite{BHfQ}.
The surface gravity for this black hole is calculated as
\begin{equation}
    \kappa_b = -\frac{\gamma  \sqrt{ C \left(\frac{r_s}{r_T}\right)^{\beta }}}{2 C r_s}.
\end{equation}

Another important thermodynamical parameter is the Hawking temperature of the black hole which is given by
\begin{equation}
    T = -\frac{\gamma  \sqrt{ C \left(\frac{r_s}{r_T}\right)^{\beta }}}{4 \pi  C r_s}.
\end{equation}

Now, if one considers $r_s = 2M$, where $M$ is the mass of the black hole, one may arrive at an expression for the entropy of the black hole given by:
\begin{equation}
    S = \frac{64 \pi  C M^2}{(\beta -4) \gamma  \sqrt{4^{\beta } C \left(\frac{M}{r_T}\right)^{\beta }}},
\end{equation}
which satisfies the first law of thermodynamics of the black hole: $dM = T dS$ with pressure $P=0$.

Following Ref. \cite{newref1, newref2}, if we consider the $rr$ components of field equations,
\begin{widetext}
\begin{equation}\label{eq55}
    8 \pi  P=\frac{r^2 f(Q)}{g(r)}-\frac{\left(\frac{1}{g(r)}-1\right) f'(Q) \left(\frac{r h'(r)}{h(r)}-\frac{r g'(r)}{g(r)}+2\right)-\frac{2 r h'(r)}{h(r)}+2 Q' r \left(\frac{1}{g(r)}-1\right) f''(Q)}{2 r^2},
\end{equation}
\end{widetext}
we can see that due to the behaviour of the $f(Q)$ gravity field equations, at horizon, several terms in the equation diverges for an ansatz $h(r) \ne g(r)$. Hence we consider the black hole solution obtained in Ref. \cite{BHfQ} and simplify the field equation.
 At the horizon $r=r_s$, we obtain,
\begin{align}
   &\left(-\frac{\beta }{r^2}\right)^{n-1} \left(2 (n-1) n r g'(r)+4 n^2-(\beta +6) n-2 \beta  r^2\right) \notag \\&+2 r g'(r)=0.
\end{align}

Using the definition of black hole temperature at the horizon, we can further write the above expression as
\begin{multline}
    \frac{8 \pi T \left(\beta +\beta  (n-1) n \left(-\frac{\beta }{r^2}\right)^{n-1}\right)}{\sqrt{C \left(\frac{r_s}{r_{T}}\right)^{\beta }}}+ \\
    r_s \left(-4 n^2+(\beta +6) n+2 \beta  r_s^2\right) \left(-\frac{\beta }{r_s^2}\right)^n=0.
\end{multline}

One may note that terms with $P$ will vanish here due to the behaviour of the black hole solution (no effective cosmological constant). We may identify the additional terms with temperature $T$ as $d\bar{S}$ in the above expression. However, one may note that $\bar{S}$ may not be precisely an entropy term here; instead, it is a normalized or scaled term mimicking the properties of entropy. Due to the property of the field equations in $f(Q)$ gravity, several terms in the denominator diverge at the horizon, so a re-scaling has been done in the limit $r \rightarrow r_s$.
From the above expression, we identify the following:
\begin{eqnarray}
    d \bar{S} &=& \frac{\left(\beta +\beta  (n-1) n \left(-\frac{\beta }{r_s^2}\right)^{n-1}\right)}{\sqrt{C \left(\frac{r_s}{r_{T}}\right)^{\beta }}},\\
    d \bar{E} &=& r_s \frac{\left(-4 n^2+(\beta +6) n+2 \beta  r_s^2\right)}{8 \pi } \left(-\frac{\beta }{r_s^2}\right)^n.
\end{eqnarray}

Here $\bar{E}$ term mimics the energy of the black hole. These expressions satisfy the first law of thermodynamics for the power law model in $f(Q)$ gravity. However, for a better understanding of the actual behaviour of these parameters, we might need to look for more viable black hole solutions in the framework of $f(Q)$ gravity. This is because, in the $f(Q)$ gravity power law model, the black hole solution considered here is not physically viable \cite{BHfQ}. Hence, to study horizon thermodynamics adequately, one needs to obtain a physically viable black hole solution at first, which is beyond the scope of this study.

In the above investigation, we have considered a theoretically motivated black hole solution of the model considered here and briefly discussed the thermodynamic variables. One may obtain the black hole temperature, surface gravity, and entropy as shown above, assuming that the first law of thermodynamics is valid. Otherwise, to realize the horizon thermodynamics, one can follow Ref.s \cite{newref1, newref2} to obtain an equivalent form of the first law from the $rr$ component of the field equations. In this case, we have considered a black hole with no effective cosmological constant resulting in the pressure $P$ associated with it being zero. This reduces the first law to the form $dE = TdS$. In our analysis, recovering $T$ from the previous definition, we only obtained an equivalent expression of the first law. To obtain an exact expression as well as to study the relevant properties, one needs to obtain a physically viable black hole solution in this theory, which we leave as a future prospect of this work. Similarly, generalized second law also may have several issues due to the presence of non-metricity and the form of field equations, and hence a detailed investigation in this regard is necessary to have a clear picture.

\section{Conclusion}

\label{sec7}

In this study, we investigated the $f(Q)$ gravity theory to examine the late cosmic expansion of the universe. A power law $f(Q)$ model, especially $f\left( Q\right) =\alpha Q^{n}$, where $\alpha $ and $n$\ are free model parameters, was taken into consideration. The field equations for the flat
FLRW geometry were then derived. Using the constant sound speed parameterizations, i.e., $\vartheta_{s}^{2}=\beta $, we can create the energy density and pressure for our $f(Q)$\ cosmic model in the current scenario, where a barotropic cosmic fluid is described in terms of $\beta $. For this model,
several cosmological parameters in terms of redshift  as well as the EoS parameter are studied. Further, we resolved the field equations using these factors and found the exact solution represented by the Hubble parameter in Eq. (\ref{H}). We were also able to determine the model parameters that fit the data sets the best using the updated Hubble data-sets (Hz), which have 31 data points, the recently published Pantheon samples (SN), which contain 1048 points, and Baryon acoustic oscillations data sets (BAO). The best-fit
values are determined using these data sets and are displayed in Table \ref{tab}. Fig. \ref{fig_q} displays the graphical behavior of the deceleration parameter $q$. Accelerated and decelerated phases are observed for $q$ in our $f(Q)$ cosmic model. According to Fig. \ref{fig_omega}, which depicts the evolution of the EoS parameter $\omega $ about redshift, $\omega $ approaches the $\Lambda $CDM point for lower values of $z $ and coincides with the quintessence epoch for larger values of $z$ in our $f(Q)$\ cosmic model. The estimated present values of the deceleration parameter corresponding to the values of the model parameters imposed by $Hz$, $SNe$, and the combined $Hz+SNe+BAO$ data sets are $q_{0}=-0.71^{+0.07}_{-0.04}$, $q_{0}=-0.66^{+0.06}_{-0.09}$, and $q_{0}=-0.91^{+0.00005}_{-0.008}$, respectively. Further, the present values of the EoS parameter are $\omega _{0}=-0.872_{-0.075}^{+0.052}$ for the Hz data-sets, $\omega _{0}=-0.868_{-0.083}^{+0.051}$ for the SN data-sets, and $\omega
_{0}=-0.9867_{-0.0084}^{+0.0054}$ for the $Hz+SNe+BAO$. Finally, we used the statefinder and $Om\left( z\right) $ diagnostics to examine how our model differed from other dark energy models. We can see from Fig. \ref{fig_rs} that the trajectory of the $r-s$ plan initially departs from the $\Lambda $%
CDM model. The $\Lambda $CDM model, which aligns with accepted cosmology, coincides with it in the late period. Also, the trajectory of the $r-q$ plan begins with the Chaplygin gas and moves on to
the de-Sitter point ($q=-1,r=1$) at the end. The $Om\left( z\right) $ diagnostic parameter, as illustrated in Fig. \ref{fig_om} represents the phantom-like era for the model. The consistency of the acquired results with accepted cosmological models and observational data sets indicates the validity of our model, making it far more attractive to scholars in this field for further research.

Finally, we investigated the horizon thermodynamics in $f(Q)$ gravity model. For this purpose, we considered a black hole solution recently obtained in Ref. \cite{BHfQ}. However, this black hole solution is not physically viable and suffers several unsolved issues. Getting a physically viable black hole solution is not directly associated with the primary objective of this study. Hence, for a qualitative analysis, we pick this solution from Ref. \cite{BHfQ} and calculate the thermodynamics parameters associated with the model. We follow Ref.s \cite{newref1, newref2} to analyze the field equations further and observe that a first law equivalent relation could be extracted from the $rr$ component of the field equation. However, the further detailed analysis would be necessary with a physically viable black hole solution to obtain an exact relation. We keep this as a prospect of the study.

\section*{Data Availability Statement}
There are no new data associated with this article.

\section*{ACKNOWLEDGMENTS}

S.A. acknowledges BITS-Pilani, Hyderabad Campus, India for an Institute fellowship. PKS acknowledges Science and Engineering Research Board, Department of Science and Technology, Government of India for financial support to carry out Research project No.: CRG/2022/001847 and IUCAA, Pune, India for providing support through the visiting Associateship program. We are very much grateful to the honorable referee and to the editor for the illuminating suggestions that have significantly improved our work in terms of research quality, and presentation.


\end{document}